\documentclass{aastex62}

\usepackage{epstopdf}
\usepackage{makecell}
\submitjournal{ApJ}
\shorttitle{Homologous Solar Eruptions From AR 11429}
\shortauthors{Dhakal et al.}
\begin{document}

\title{Recurring Homologous Solar Eruptions in NOAA AR 11429}

\correspondingauthor{Suman Dhakal}
\email{sdhakal2@gmu.edu}

\author{Suman K. Dhakal}

\author{Jie Zhang}
\affiliation{Department of Physics and Astronomy, George Mason University, 4400 University Dr., MSN 3F3, Fairfax, VA 22030, USA}
\

\author{Panditi Vemareddy}
\affiliation{Indian Institute of Astrophysics, II Block, Koramangala, Bengalure-560 034, India}

\author{Nishu Karna}
\affiliation{Harvard-Smithsonian Center for Astrophysics, Cambridge, MA, 02138, USA}

\begin{abstract}
We present the study of three homologous solar eruptions from NOAA active region (AR) 11429 over four days. This large and complex AR divided into two relatively simple sub-regions: northeast (NE) and southwest (SW). Recurrent eruptions occurred from the SW sub-region over different evolutionary phases, which provided a unique opportunity to isolate the physical processes responsible for solar eruptions. Persistent shearing and convergence of opposite magnetic polarities led to continuous flux cancellation along the SW polarity inversion line (PIL). A filament persistently lying along the SW-PIL was observed to survive each eruption, which suggests the partial eruption of the magnetic system. Further, following the first and second eruptions, a sigmoidal magnetic structure of similar morphology was reformed along the SW-PIL. The photospheric motion of magnetic flux continuously injected and stored the negative helicity in the partially erupted magnetic system and built up the magnetic free energy for the successive eruptions. These results suggest that the shearing motion and magnetic flux cancellation of opposite fluxes were: (1) the dominant factor, irrespective of the evolutionary phase, that contributed to the recurrent homologous eruption, and (2) the key processes of forming the erupting structure, likely a magnetic flux rope, and its long-lasting continuation results in reformation of identical erupting structure. The study also finds that similar magnetic topology could result in the magnetic reconnection at the same location, and such flares during the precursor phase would help in the eruption by decreasing the constraint of overlying magnetic field.

\end{abstract}


\section{Introduction} 
\label{sec:intro}
Solar eruptions, manifested as flares and Coronal mass ejections (CMEs), are the most spectacular phenomena happening on the solar corona. Besides being magnificent, these are the main sources of the disturbances in the interplanetary space and also the space weather effects near the Earth. Flares and CMEs (often occur together) are probably the manifestation of a single energy release process, during which a tremendous amount of stored magnetic energy is released from the solar atmosphere (see, \citealt{Forbes_etal_2000}; \citealt{Zhang_etal_2001}). The general model of solar eruptions, also known as CSHKP model (initially developed by \citealt{Carmichael_1964}; \citealt{Sturrock_1966}; \citealt{Hirayama_1974}; \citealt{Kopp_and_Pneuman_1976}), describes the process as the eruption of magnetic flux rope (MFR) through magnetic reconnection. An MFR is a highly sheared and twisted magnetic field structure and carries a large amount of free magnetic energy. It is kept under equilibrium by the magnetic pressure of the overlying magnetic fields and its instability makes it to rise. While rising it drags the overlying magnetic field lines, making them to come close to each other in an anti-parallel manner below it and form a current sheet. The magnetic reconnection at current sheet results in flares and the new connection removes the overlying constrain over MFR and accelerate it (see, \citealt{Forbes_etal_2000}; \citealt{Chen_2011} for review). As long as there is an erupting MFR, the general theory describes the evolution of solar eruptions very well. However, it does not address the initial formation of MFRs and onset of solar eruptions.

Many agree with the notion that the MFR is an essential feature for solar eruption. However, there is a debate regarding when, how and where it is formed. It can exist before (e.g., \citealt{Kliem_and_Torok_2006}; \citealt{Zhang_etal_2012}) or form during solar eruption (e.g., \citealt{Antiochos_etal_1999}; \citealt{Cheng_etal_2011}). Its detection before the eruption is a difficult task, due to low plasma density and technological insufficiency to measure magnetic field directly in the corona. Nevertheless, its existence is suggested through sigmoidal bright loops in X-rays (\citealt{Rust_and_Kumar_1996}; \citealt{Green_etal_2007}), AIA/SDO hot-channel structures (\citealt{Cheng_etal_2011}; \citealt{Zhang_etal_2012}), dips or bald patches in filament channels (\citealt{Lites_etal_2005}; \citealt{Lopez_Ariste_etal_2006}), and nonlinear force-free field (NLFFF) extrapolation results (e.g., \citealt{Chintzoglou_etal_2015}). Several mechanisms are associated with its formation, such as flux emergence (e.g. \citealt{Fan_and_Gibson_2003}; \citealt{Leake_etal_I_2013}), shearing motion between opposite magnetic fluxes (e.g., \citealt{Amari_etal_2000,Amari_etal_I_2003, Jacobs_etal_2009}), magnetic flux cancellation along the polarity inversion lines (PILs) (e.g., \citealt{van_Ballegooijen_etal_1989, Aulanier_etal_2010, Green_etal_2011}), and confined flaring (e.g., \citealt{Patsourakos_etal_2013}, \citealt{Chintzoglou_etal_2015}). 

A typical solar eruption has three phases: (1) the precursor phase (PP), (2) the impulsive phase (IP), and (3) the gradual phase (GP; see, \citealt{Zhou_etal_2016} for discussion); these are associated with three distinct acceleration phases of CMEs (\citealt{Zhang_etal_2001}). PP is important to understand the initiation and validating different models of solar eruptions. In the past, different observational signatures have been identified in precursor phase for certain individual events, such as: soft X-rays (SXR) emission (\citealt{Farnik_etal_1996}; \citealt{Farnik_and_Savy_1998}), hard X-rays burst (\citealt{Harra_etal_2001}), UV/EUV brightenings (e.g., \citealt{Joshi_etal_2011}; \citealt{Awasthi_etal_2014}), type II radio bursts (e.g., \citealt{Klassen_etal_2003}; \citealt{Liu_etal_2007}), prominence oscillation (\citealt{Chen_etal_2008}) and MFR oscillation (\citealt{Zhou_etal_2016}). However, due to weak or entirely absent observational activities during precursor phase its current understanding is very poor compared to IP and GP (\citealt{Hudson_2011}). 

Most of the time, single and independent solar eruption happens on the Sun. However, there are cases where multiple eruptions occur in a short time-span. Such eruptions are known as sympathetic eruptions. These could occur in different active regions (ARs) or at different locations within the same complex AR. Although interrelation among these is a debatable issue (see, \citealt{Biesecker_and_Thompson_2000}), but causal relationship have been identified in many cases (e.g., \citealt{Schrijver_and_Title_2011}; \citealt{Torok_etal_2011}; \citealt{Liu_etal_II_2009}). 

Besides sympathetic eruptions, multiple eruptions could happen from the same local region consecutively. Consecutive CMEs originating from the same region, associated with homologous flares, having similar coronographic appearance and similar dimming in extreme ultraviolet imaging are known as homologous CMEs \citep{Zhang_and_Wang_2002}. Homologous eruptions are possible only if there are mechanisms that continuously store magnetic energy in the same local region. Therefore, their studies are important to understand the long-term magnetic energy build-up in the corona. In the past, they have been studied either during magnetic flux emergence phase and were considered due to flux emergence (e.g. \citealt{Nitta_and_Hudson_2001}; \citealt{Chatterjee_and_Fan_2013}) or flux decaying phase and were considered due to shearing motion and magnetic flux cancellation (e.g., \citealt{Li_etal_2010}; \citealt{Vemareddy_2017}). The persistent photospheric horizontal motion of magnetic structure along the PIL was also considered to produce homologous eruptions (e.g., \citealt{Romano_etal_2015}; \citealt{Romano_etal_2018}). Past studies also suggest that homologous eruptions can be triggered by similar mechanisms, such as moving magnetic features (\citealt{Zhang_and_Wang_2002}), coronal null point magnetic configuration (e.g., \citealt{DeVore_and_Antiochos_2008}) and shearing motion and magnetic reconnection (e.g., \citealt{Vemareddy_2017})

In this paper, we analyze three homologous eruptions, viz., SOL2012-03-07T01:05, SOL2012-03-09T03:22, and SOL2012-03-10T16:50. These eruptions originated from the same location in the complex AR 11429, but at different magnetic flux evolutionary phase. An MFR of similar morphology was formed, along the same PIL, before each of the three eruptions. Magnetic flux cancellation led by the shearing and converging motion and new connectivity in the corona were responsible for its formation. A confined flare was observed during the PP of each eruption, which helped in the eruption by weakening the constraint of overlying magnetic fields. The paper is structured as follows. Instrument and data are described in Section \ref{instru}. The precursor phase and each eruption are described in Section \ref{eruption}. Evolution of the AR before each eruption and pre-eruptive coronal magnetic structures are described in Section \ref{obs}, and the discussion and conclusion are presented in Section \ref{DC}.

\section{\textbf{Instruments}}
\label{instru}
This study mainly used the data from the Atmospheric Imaging Assembly (AIA; \citealt{Lemen_etal_2012}) and Helioseismic and Magnetic Imager (HMI; \citealt{Schou_etal_2012}), onboard the \textit{Solar Dynamics Observatory} (\textit{SDO} ;~\citealt{Pesnell_etal_2012}). The AIA provides full-disk images of the Sun with unprecedented temporal cadence (12\,s) and spatial resolution ($1\farcs2$). The different spectral bands of AIA observe the Sun from the photosphere, chromosphere, and corona up to 0.5 $R_{\sun}$ above the solar limb. The HMI provides the photospheic line-of-sight and vector magnetic field observations at a cadence of 45\,s and 720\,s respectively. The SDO has an inclined geosynchronous orbit around the Earth. During two seasons of each year, the view of the \textit{SDO} gets obscured by the Earth temporarily each day for two-three weeks (SDO eclipse;~\citealt{Pesnell_etal_2012}). In this study, we focused on the \textit{SDO} observation during the solar front disk passage of the NOAA AR 11429 from 2012 March 6,  00:00 UT to 2012 March 10, 19:00 UT. There were some data-gaps during the observational period due to SDO eclipse. In addition, we used observations from the Large Angle Spectrometric Coronagraph (LASCO; \citealt{Brueckner_etal_1995}) on board the \textit{Solar and Heliospheric Observatory} (\textit{SOHO}), white light images of the Sun-Earth Connection Coronal and Heliospheric Investigation (SECCHI; \citealt{Howard_etal_2008}) onboard \textit{Solar Terrestrial Relations Observatory} (\textit{STEREO}; \citealt{Kaiser_etal_2008}), H$\alpha$ image from $\textit{Big Bear Solar Observatory}$ ($\textit{BBSO}$;~\citealt{Denker_etal_1999}), X-ray Telescope (XRT;~\citealt{Golub_etal_2007}) onboard the \textit{Hinode} (\citealt{Kosugi_etal_2007}), and the Solar X-ray Imager (SXI;~\citealt{Hill_etal_2005} and \citealt{Pizzo_etal_2005}) on board the \textit{Geostationary Operational Environmental Satellite} ($\textit{GOES)}$.

\section{\textbf{Observations of the Homologous Eruptions}}
\label{eruption}
{\bf The} NOAA AR 11429 was initially emerged on the backside and appeared on the eastern limb of the Sun on 2012 March 4. During its front disk transition, the AR acquired very complex magnetic-configuration ($\beta/\gamma/\delta$), with many small magnetic polarities (see \citealt{Elmhamdi_etal_2014} for more details about the evolution of magnetic configuration).  However, as discussed in~\citealt{Chintzoglou_etal_2015} and~\citealt{Dhakal_etal_2018}, it could be divided into two relatively simpler sub-regions namely: northeast (NE) and southwest (SW) sub-regions, more discussed in the section~\ref{photo_evo}. Also, it had an anti-hale magnetic configuration i.e., the leading magnetic polarity (positive) was opposite of the normal hemispheric trend (negative). Both $\delta$ and anti-hale magnetic configuration are identifiable with severe flaring activity (e.g., \citealt{Zirin_1970}; \citealt{Tanaka_1991}; \citealt{Sammis_etal_2000}). Indeed, it was super-flare productive. Most of the flares were observed before it crossed the central meridian on March 9. By the end of March 9, it had produced 3 X-class, 11 M-class, and 30 C-class flares. Afterward, there were only 2 M-class flares and 7 C-class flares. The events under study occurred in the SW sub-region in the interval of $\sim$4 days. These homologous eruptions are summarized in Table \ref{table:summary}.

\begin{figure*}[!ht]
\centering
\includegraphics[width=.99\textwidth,clip=]{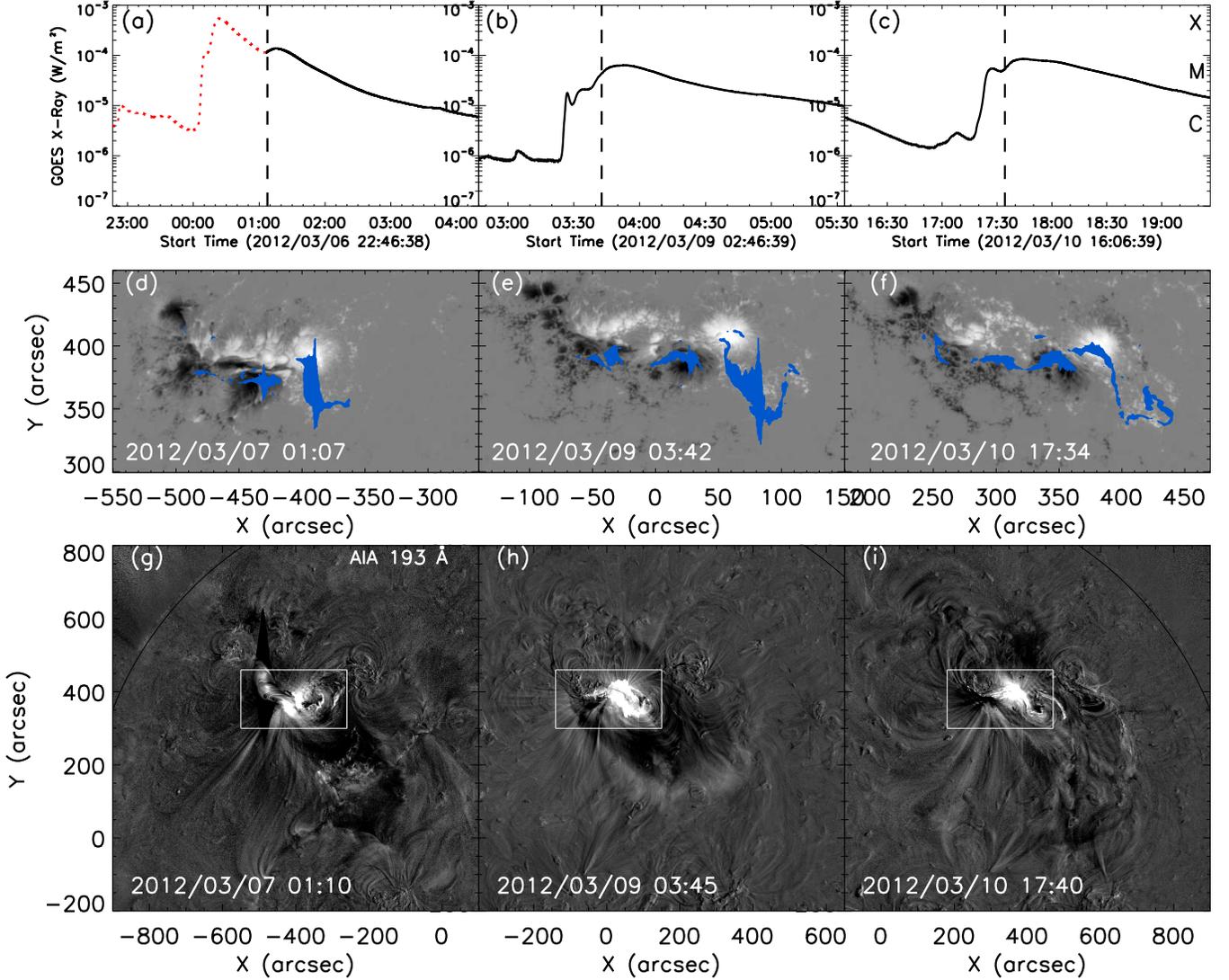}
\caption{Three homologous eruptions from NOAA AR 11429. (a-c) time profiles of GOES soft X-ray intensity flux during the three eruptions. The maximum flux in these profiles corresponds to X1.3, M6.3, M8.4 flares, respectively. The red dotted line in (a) shows the contamination in the X-ray profile due to an earlier eruption. (d-f) Overlay of the flare ribbons (in blue) observed in AIA 1600 \AA~on HMI LOS-magnetogram. (g-i) Difference images of AIA 193~\AA~ showing the EUV dimming associated with each eruption. The white box represents the FOV of the magnetograms shown in (d-f).}
\label{fig:dimming}
\end{figure*}

\begin{figure*}[!ht]
\centering
\includegraphics[width=.7\textwidth,clip=]{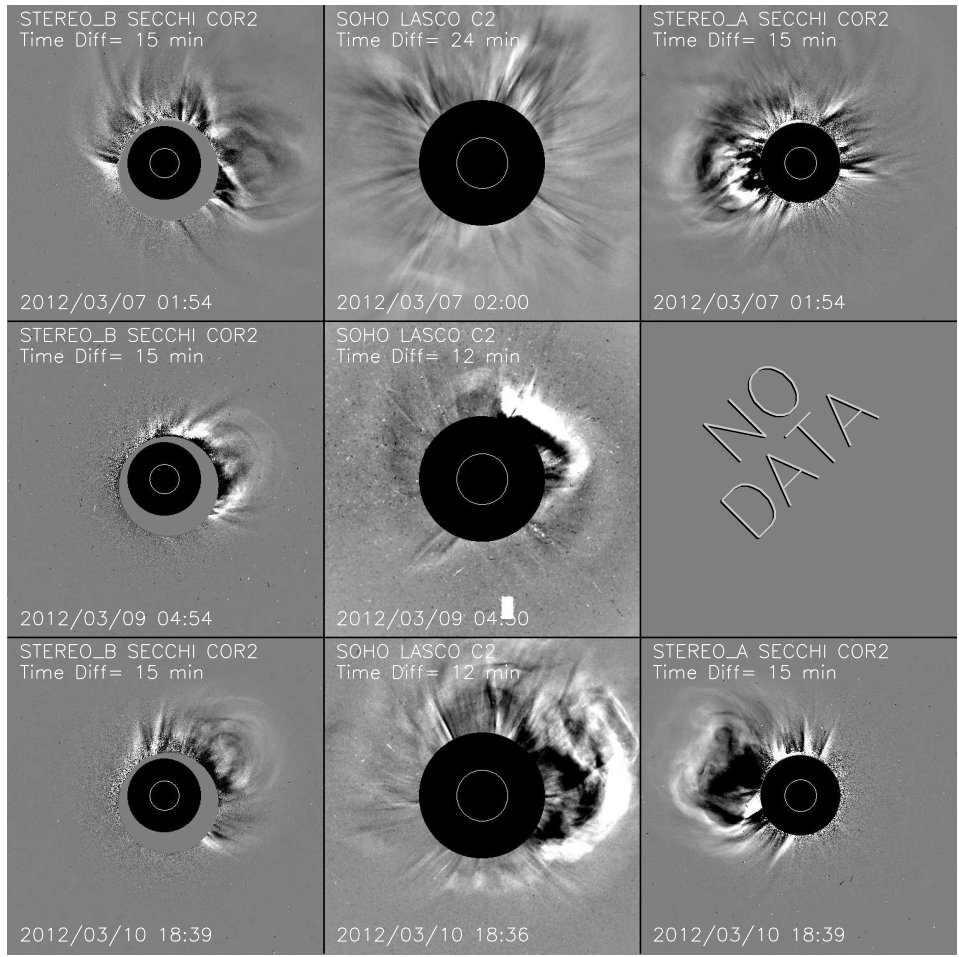}
\caption{Homologous CMEs seen from three different view points. Left panels: the COR2 images from SECCHI/STEREO-B. Middle panels: the C2 images from LASCO/SOHO. Right panels: the COR2 images from SECCHI/STEREO-A.}
\label{fig:corona}
\end{figure*}

\begin{table*}[!ht]
\centering
\begin{tabular}{ |c|c|c|c|c|c| } 
 \hline
 Events & Initiation Time$^{a}$ & Location$^{b}$ & \makecell{Associated Flare\\ (Peak Time)$^{c}$} & \makecell{LASCO C2 \\Appearance$^{d}$} & \makecell{LASCO Speed \\(km/s)$^{e}$} \\ 
 \hline
 CME1 & 2012/03/07 01:05 & N15 E26 & X1.3 (01:14 UT) & 2012/03/07 01:30 UT & 1825 \\ 
 \hline
 CME2 & 2012/03/09 03:22 & N15 W03 & M6.3 (03:53 UT) & 2012/03/09 04:26 UT & 950 \\ 
 \hline
 CME3 & 2012/03/10 16:50 & N17 W24 & M8.4 (17:44 UT) & 2012/03/10 18:00 UT & 1296 \\
 \hline
 \end{tabular}
 \caption{Three homologous solar eruptions from AR 11429.\\
    $^{a}$Initiation time of the solar eruption\\
    $^{b}$flare location\\
    $^{c}$peak time of GOES X-ray flux\\
    $^{d}$first LASCO C2 appearance\\
    $^{e}$linear speed of CME in LASCO C2, obtained from \url{https://cdaw.gsfc.nasa.gov/CME_list/UNIVERSAL/2012_03/univ2012_03.html}}
    \label{table:summary}
\end{table*}
The first solar eruption occurred on 2012 March 7 at $\sim$01:05 UT when the AR was in the north-east quadrant of the solar disk. On March 7, within one hour, there were two solar eruptions viz., SOL2012-03-07T00:02 (shown in red dotted line in Figure \ref{fig:dimming}(a)) and SOL2012-03-07T01:05 (shown in black line in Figure \ref{fig:dimming}(a)) from the NE and SW sub-regions respectively (see \citealt{Chintzoglou_etal_2015}). Here, we study the eruption from the SW sub-region. It was associated with X1.3 class flare and its location was $\sim$N15E26. Associated flare ribbons were concentrated mainly on the positive and negative fluxes of the SW sub-region (see Figure \ref{fig:dimming}(d) for observed flare ribbon in AIA 1600 \AA) and were observed to move away on either side of the PIL during the MFR eruption. The position and separation of the flare ribbons indicated that the MFR erupted from the SW-PIL. The eruption led to the coronal mass depletion in the region and observed as coronal dimming in different AIA passbands (e.g. \citealt{Tian_etal_2012}; see Figure \ref{fig:dimming}(g) for observed dimming in AIA 193 \AA). Later it was observed as halo CME, moving with 1825 km/s (CDAW CME catalog), in $\textit{LASCO}$ white light observation at 01:30 UT. The morphology of the CME at a single time snapshot is not clear due to contamination from the earlier CME. However, its morphology was clear in $\textit{STEREO-A}$ FOV as eastward-moving CME and in $\textit{STEREO-B}$ FOV as westward-moving CME (see upper panels of the Figure~\ref{fig:corona}).

The second solar eruption started on 2012 March 9, at $\sim$03:22 UT when the AR was located around the central meridian of the solar disk. It was associated with an M6.3 class flare (see Figure \ref{fig:dimming}(b)) at $\sim$N15W03. The flare ribbons, similar to the ribbons of the previous eruption, evolved and concentrated on the opposite magnetic flux in the SW sub-region (see Figure\ref{fig:dimming}(e)). This again indicated that the MFR had erupted from the SW-PIL. A dimming similar to the previous eruption was observed in EUV observation (see Figure \ref{fig:dimming}(h)) and eventually observed as halo CME in $\textit{LASCO}$ C2 at 04:26 UT. In the $\textit{LASCO}$ FOV CME moved with a linear velocity of 950 km/s. $\textit{STEREO-A}$ data was not available during that period, but like the earlier eruption it appeared as westward-moving CME in $\textit{STEREO-B}$ FOV (see middle panels of the Figure~\ref{fig:corona}). 

The third solar eruption started on 2012 March 10, at $\sim$ 16:50 UT when the AR was at the north-west quadrant. It was associated with an M8.4 class flare (see Figure \ref{fig:dimming}(c)) at $\sim$N17W24. The evolution and separation of flare ribbons were similar to the previous two eruptions and they were primarily concentrated on the opposite magnetic flux of the SW sub-region (see Figure\ref{fig:dimming}(f)). Therefore, the third MFR also erupted from the SW-PIL. Observed EUV dimming was also similar to the previous two eruptions (see Figure \ref{fig:dimming}(i)). It appeared as halo CME in $\textit{LASCO}$ C2 at 18:00 UT, moving with a linear velocity of 1296 km/s. Like earlier eruptions, it was observed as eastward-moving CME in $\textit{STEREO-A}$, and westward-moving CME in $\textit{STEREO-B}$ (see the lower panels of Figure~\ref{fig:corona}) with a similar CME morphology.

All three solar eruptions were originated from the SW-PIL of the AR 11429. The direction of successive CMEs seemed to move towards north-side in $\textit{STEREO}$ FOV and this could be due to the change in magnetic flux distribution with evolution. Nevertheless, the overall morphology of the CMEs appeared similar in each recurring eruptions. Also, the X-ray profile, evolution of flare ribbons, and EUV dimming associated with these eruptions were nearly identical to be homologous eruptions. The time difference between the first and the second eruption was $\sim$50 hours, and the time difference between the second and the third eruption was $\sim$37 hours. The next section discusses the evolution of the AR that led to identical solar eruptions. 

\section{\textbf{Evolution of AR 11429 leading to homologous eruptions}}
\label{obs}
\begin{figure}[!ht]
\centering
\includegraphics[width=.5\textwidth,clip=]{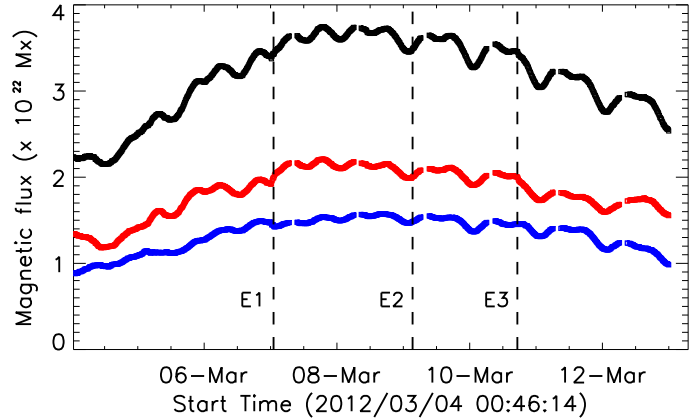}
\caption{Time evolution of the magnetic-flux in the AR 11429 during March 4-13. The total unsigned magnetic flux is plotted in black, the total negative magnetic flux is plotted in red and the total positive flux is plotted in blue. Dashed vertical lines marks the time of three homologous solar eruptions.}
\label{fig:mag_flux}
\end{figure}

We used the preprocessed cutouts of SHARP data products (\texttt{hmi.sharp\_cea\_720s}) to study the photospheric evolution of the magnetic flux distribution in the AR. Preprocessing resolves the 180$^\circ$ azimuthal uncertainty and remaps helioprojective images into a cylindrical equal-area (CEA) projection, where each pixel has the same surface area (\citealt{Hoeksema_etal_2014}). Although there were several episodes of new magnetic flux emergence in the early evolutionary period, no major emergence was observed after March 8, as is evident from the time-profile of magnetic flux of the AR shown in the Figure~\ref{fig:mag_flux} (also, see the movie accompanying Figure~\ref{fig:plasma_velocity}). Here, we used the CEA maps of HMI LOS-magnetogram to analyze the magnetic flux. The apparent 12-hour periodicity in the plots of magnetic flux (in Figure~\ref{fig:mag_flux}) is due to instrumental  effect and related to spacecraft orbital velocity, also there is a variation of the number of pixels contributing to the low-to-moderate magnetic field values as AR moves from central meridian to the limb (see, \citealt{Hoeksema_etal_2014}). Solar eruptions under consideration had occurred at the different evolutionary phases of the AR (eruptions are shown by dashed-vertical lines in Figure~\ref{fig:mag_flux}). The first eruption had occurred when the total magnetic flux was increasing i.e. during the magnetic flux emergence phase. The second eruption happened at the end of the flux emergence phase, and the third eruption happened during the flux decay phase of the AR. In the following, we describe the evolution of the AR, focusing on SW sub-region, in the period of 2012 March 6 to March 11. 
 
\subsection{\textbf{Evolution of the Photospheric Magnetic Field}}
\label{photo_evo}

\begin{figure*}[h]
\centering
\includegraphics[width=.99\textwidth,clip=]{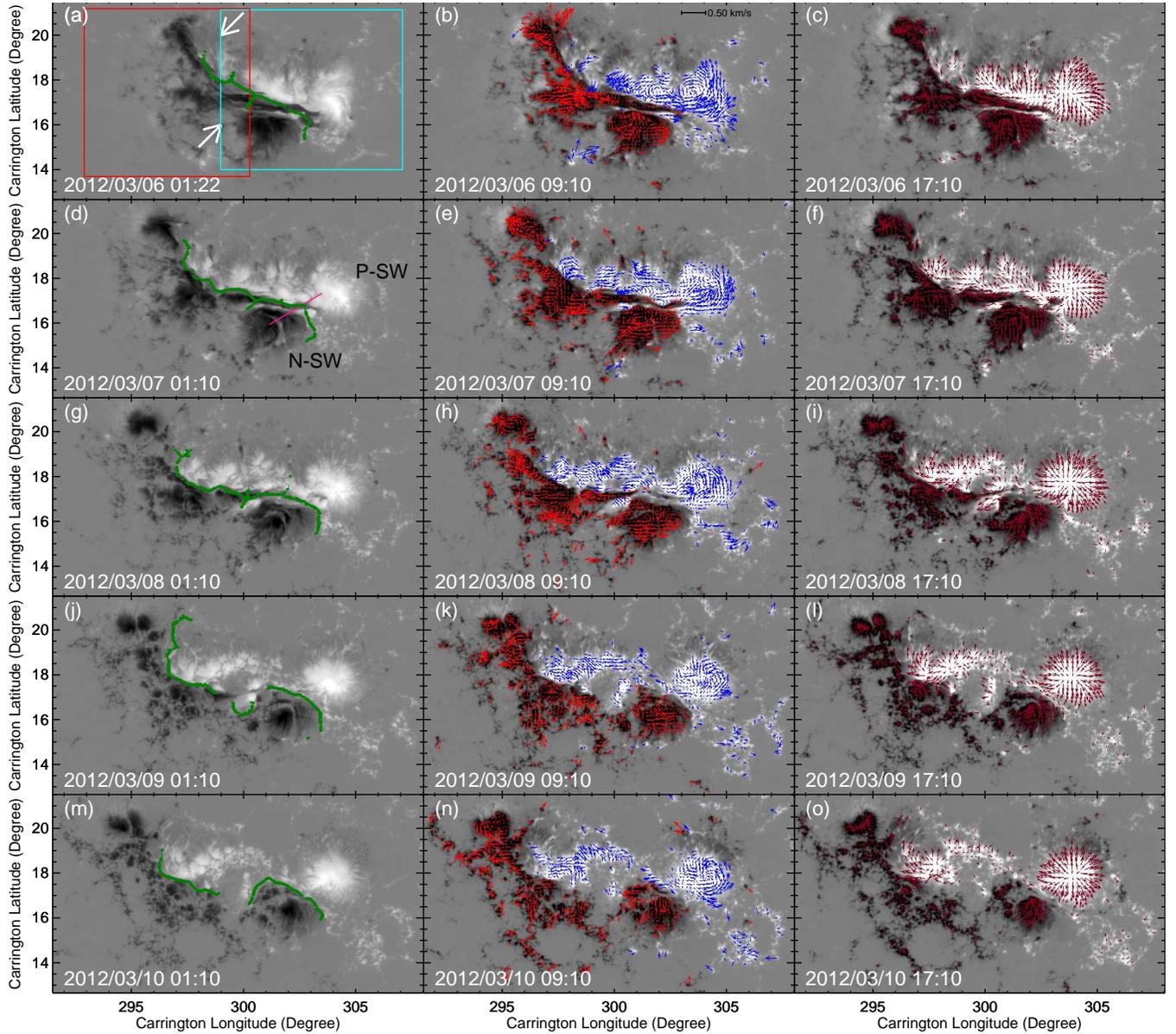}
\caption{Evolution of the photospheric magnetic field in NOAA AR 11429. Images are taken from the line-of-sight magnetograms of SDO/HMI and shown in a cylindrical equal-area heliographic projection. Left panels: the strong-gradient polarity inversion lines in green asterisks are overplotted on the HMI LOS-magnetogram. The red and cyan boxes, in (a), represent the northeast (NE) and southwest (SW) sub-regions respectively, and white arrows show the location of new flux emergence. The pink line in (d) shows the distance between the centroid of positive (P-SW) and negative (N-SW) polarity in the SW sub-region. Middle panels: the direction of motion of positive (negative) flux are overplotted on the HMI LOS-magnetogram in blue (red) arrows. An animation of these panels is available starting on 2012 March 6, 00:10 until 2012 March 10, 19:46 UT. The video duration is 22 s.  Right panels: the direction of the horizontal component of magnetic fields are overplotted on the HMI LOS-magnetogram in brown arrows. (An animation of this figure is available at \url{http://solar.gmu.edu/tmp/homologous_eruption/velocity.mpg}.)}
\label{fig:plasma_velocity}
\end{figure*}

\begin{figure*}[!ht]
\centering
\includegraphics[width=.8\textwidth,clip=]{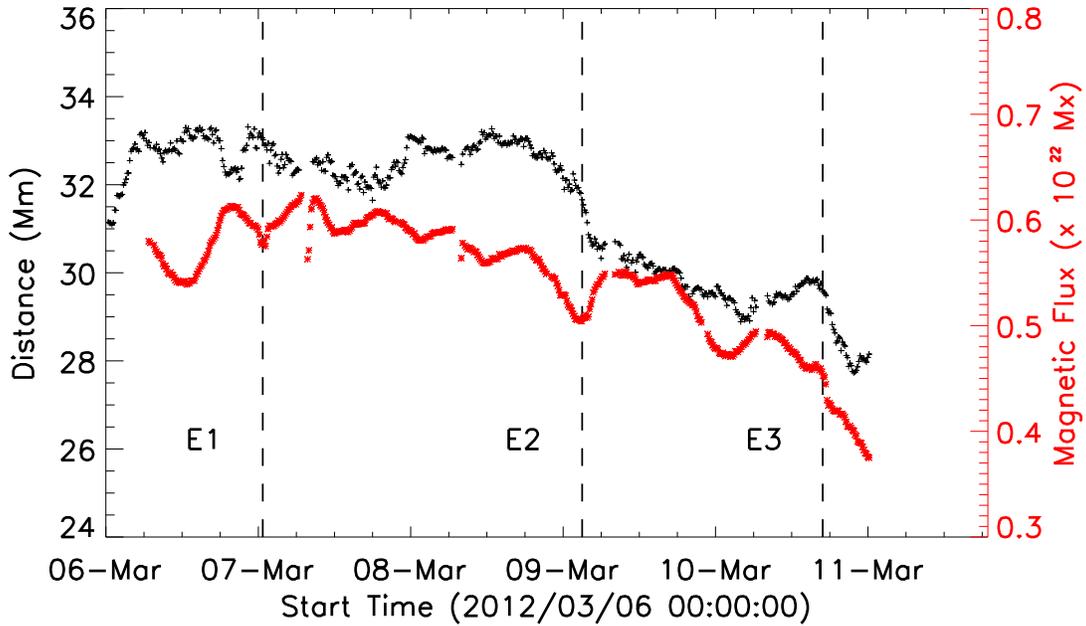}
\caption{Convergence of opposite magnetic polarity and magnetic flux cancellation in the southwest sub-region of the AR 11429. Convergence rate is examined by monitoring the distance between the centroid of positive and negative magnetic polarity and is shown in black asterisks. The time profile of negative flux in the SW sub-region is in red. The vertical dashed lines refer to the time of three solar eruptions from the AR.}
\label{fig:converge_plot}
\end{figure*}

 Initial compactness and complexity of the magnetic flux distribution of the AR decreased with its evolution. It was divided into two simpler sub-regions: NE and SW based on the presence of two erupting MFRs on 2012 March 7 (\citealt{Chintzoglou_etal_2015}) and are shown in red and cyan box respectively in Figure~\ref{fig:plasma_velocity}(a). The distinction between the two sub-regions become more clear in the later evolutionary period. Long and sharp strong-gradient PILs, shown in green asterisks in the left panels of Figure~\ref{fig:plasma_velocity} (obtained from an image gradient operation as discussed in~\citealt{Zhang_etal_2010}), were present throughout the observational period. The plasma velocity maps show the persistent shearing motion between opposite magnetic fluxes along the SW-PILs (see the middle panels of Figure~\ref{fig:plasma_velocity}; also see the movie accompanying Figure \ref{fig:plasma_velocity} for the plasma velocity). These maps were derived using DAVE4VM method with an apodization window of 19 pixels (\citealt{Schuck_2008}). Shearing motions along the PILs are well known to convert potential field to sheared field arcade (e.g., \citealt{van_Ballegooijen_etal_1989}). The direction of horizontal magnetic fields, which were aligned along the length of PILs (see the right panels of Figure \ref{fig:plasma_velocity}), indicated that there was indeed strong non-potential shear along the PILs. The SW-PIL was lying between two large areas of opposite magnetic polarity i.e., P-SW and N-SW (see Figure~\ref{fig:plasma_velocity}(d)). The distance between the centroid of opposite magnetic poles was decreasing with time (see the plot in black-asterisks in Figure~\ref{fig:converge_plot}). This showed that the opposite magnetic poles were continuously converging towards each other. The centroid-distance was decreased by $\sim$1 Mm between first and second solar eruptions (from March 7 to March 9) and by $\sim$2 Mm between second and third eruptions (from March 9 to March 10). Though the convergence was observed in both flux emerging and decaying phase, the rate of convergence was increased after the flux emergence had been stopped. Flux cancellation was not obvious from the plot of the magnetic flux of the entire AR and it was hard to isolate the SW sub-region completely. However, after March 6 it was possible to isolate the negative polarity of the SW sub-region. Therefore we encompassed the negative polarity with an ellipse and ellipse was varying with time to make sure that it fully encompassed the negative polarity all the time, and analyzed it. The plot in red asterisks in Figure~\ref{fig:converge_plot} shows the time-profile of negative flux in the SW sub-region. The negative flux was decreasing gradually, indicating a continuous flux cancellation in the SW sub-region. Though the rate of convergence was increased after the second eruption on March 9, there was no significant change in the slope of the negative magnetic flux, which indicates that the flux cancellation rate was almost the same during and after the flux emergence phase. 

\begin{figure}[!ht]
\centering
\includegraphics[width=.65\textwidth,clip=]{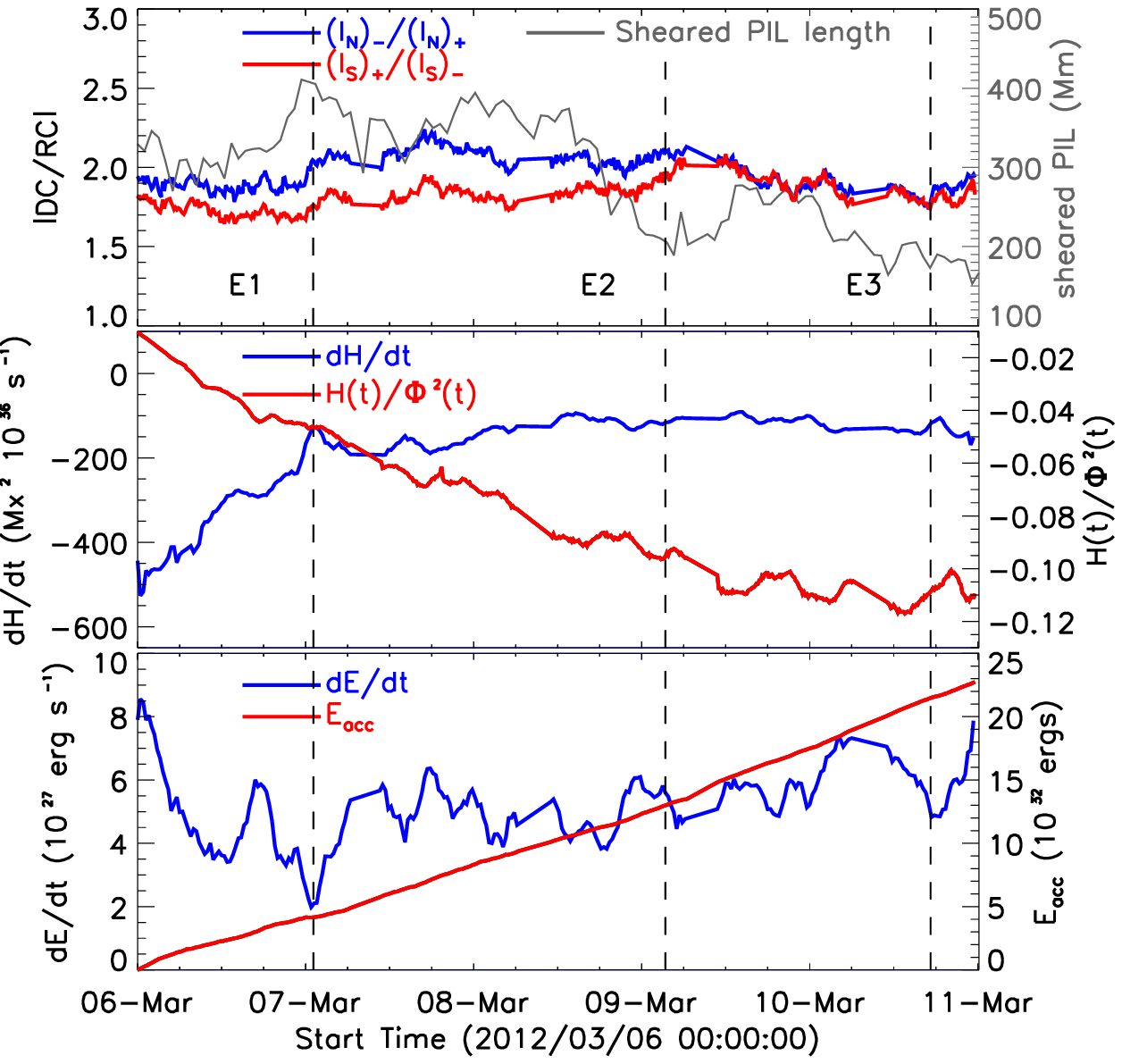}
\caption{Evolution of non-potential parameters in AR 11429. Top panel: Degree of net current neutralization in positive (negative) polarity in blue (red). A higher value of around 1.8 suggest non-neutralized currents in the AR flux system and instability. The length of sheared PILs is also plotted in grey. Middle panel: the time rate of helicity flux is plotted in blue. Normalized accumulated helicity (time-integrated helicity flux normalized with square of the flux) is plotted in red. Bottom panel: energy flux injection (poynting flux) and its accumulated quantity in blue and red respectively.}
\label{fig:poynting}
\end{figure}

To understand the high eruptivity of the AR 11429, we analyzed certain non-potential parameters: viz, non-neutralized current, length of sheared PIL, helicity ($dH/dt$), and energy injection ($dE/dt$) rate (for computational details of these parameters see, \citealt{Vemareddy_2015}; \citealt{Vasantharaju_2018_non_pot}; \citealt{Vemareddy_2019_apj}). The upper panel of Figure~\ref{fig:poynting} shows the non-neutralized currents in positive (negative) polarity in blue (red). A strong non-neutralized current was observed in the AR 11429 throughout the observational period. Also, long sheared PILs were observed throughout the evolution of the AR (shown in grey in the upper panel of Figure~\ref{fig:poynting}). The stress in the field lines due to plasma flows generates non-neutralized current in an AR (e.g., \citealt{Torok_2014_EleCurr,Vemareddy_2015_flx_emer,Vemaredd_2017_field_strength_gradient}). A higher level of deviation ($>$ 1.5) from the current neutrality condition and long sheared PILs suggests that the AR was very CME productive (e.g., \citealt{Sunx_2017_ele_neut,Vemareddy_2019}). There was a significant decrease in the length of sheared PIL after March 9, and maybe that was the reason for the decrease in the flare activity after it. Further, continuous shearing and converging motions of opposite magnetic fluxes were injecting and storing magnetic helicity and magnetic energy in the solar corona (see the middle and lower panels of Figure~\ref{fig:poynting}). The rate of helicity and energy injection was almost constant after March 7. The initial high helicity and energy rates were due to major magnetic flux emergence during the early evolutionary phase of the AR. Continuous shearing and converging motion of the opposite magnetic fluxes were continuously injecting magnetic helicity and energy in the AR 11429 and were responsible for recurrent solar eruptions. The homologous eruptions can be understood as a cycle of storage and release of magnetic helicity and energy from the AR \citep{Vemareddy_2017}.

\subsection{\textbf{Pre-eruptive Structures in the Corona}}
While strongly influenced by magnetic evolution in the photosphere, solar eruptions occur in the corona. Therefore, coronal magnetic structures and their evolution are one of the most important aspects of solar eruptions. One way to analyze them is through the observation of plasma-emission at different electromagnetic wavelengths. Another way to know about the coronal magnetic field is through the extrapolation of photospheric magnetic field data. However it is not possible to get the exact 3-D magnetic field in the corona through 2-D magnetic flux observations. Therefore observational validation is necessary for the analysis of such results. Here, we used both techniques to complement each other and to have a proper understanding of pre-eruptive coronal structures and environment. To reconstruct the coronal magnetic structure, we carried out NLFFF extrapolation (\citealt{Wiegelmann_and_Inhester_2010}) at every hour from 2012 March 6, 00:10 UT to 2012 March 10, 19:34 UT. Cutouts of HMI vector magnetograms were taken as the bottom boundary of the computational domain. These were inserted in an extended field of view to weaken the effects of lateral boundaries and were pre-processed to make them suitable for the force-free conditions (\citealt{Wiegelmann_etal_2006}). Computations were performed on a uniformly spaced grid of $276 \times 148 \times 196$ pixels corresponding to a physical volume of $199 \times 106 \times 141$ Mm$^{3}$ (more details in \citealt{Dhakal_etal_2018}).

\begin{figure*}[!ht]
\centering
\includegraphics[width=.9\textwidth,clip=]{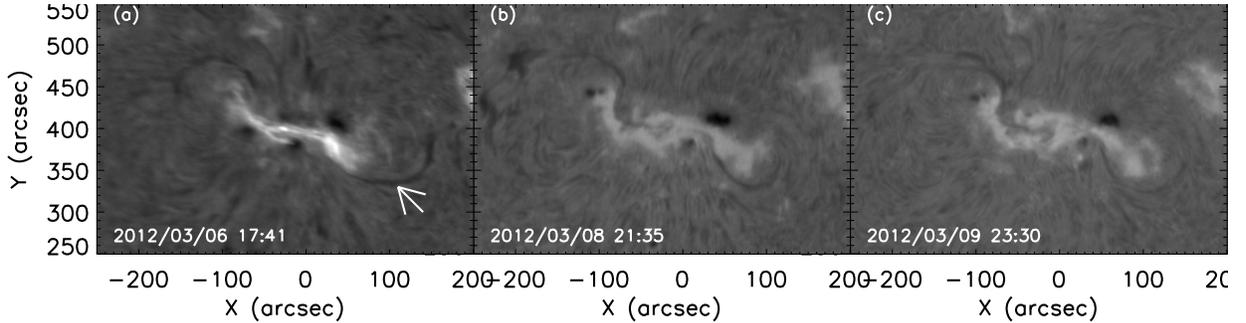}
\caption{Filament of identical shape, in H$\alpha$ images from BBSO, were present persistently during the evolution of the AR 11429. White arrow in (a) {points to} the filament in southwest sub-region of the AR.}
\label{fig:h_alpha_map}
\end{figure*}

A filament was observed to lie along the SW-PIL (see Figure~\ref{fig:h_alpha_map}), indicating sheared and/or twisted field lines (e.g., \citealt{Priest_etal_1989}). Following the eruptions on March 7 and 9, the filament was observed to still exist along the SW-PIL under the fading post-flare arcade in AIA cool passbands (e.g., 304 \AA). The appearance of the filament, immediately after the eruption, suggests that either only part of it erupted or it did not erupt at all, thus indicating the partial eruption of the magnetic structure (e.g., \citealt{Tang_1986}; \citealt{Gilbert_etal_2000}; \citealt{Gibson_etal_2002}).  

\begin{figure*}[!ht]
\centering
\includegraphics[width=.9\textwidth,clip=]{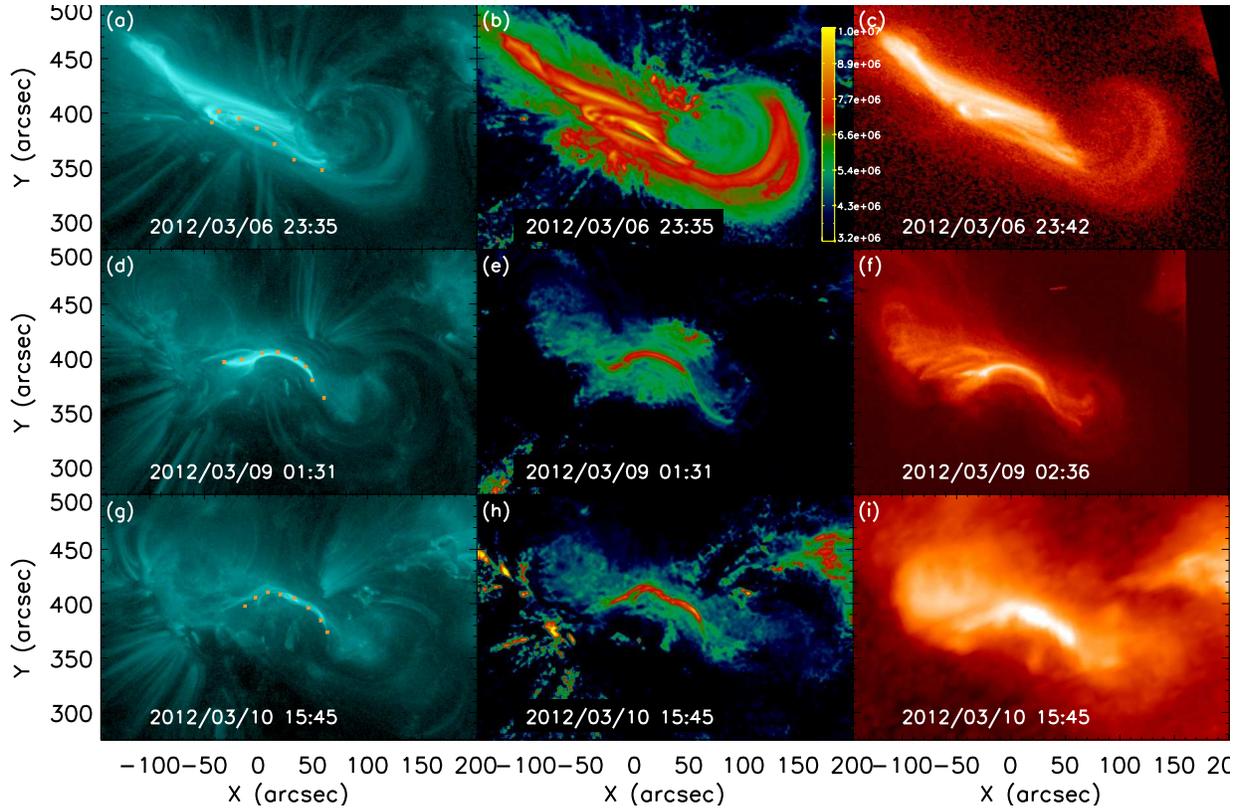}
\caption{Observation of pre-eruptive coronal structure. Left panels: the hot-channel structures (HCS) before each eruptions in AIA 131 \AA. The brighter part of the HCS is outlined by orange asterisk. Middle panels: the temperature map obtained using the DEM method. Right panels: the sigmoidal hot coronal structures in X-ray images. (c) and (f) images are from $\textit{Hinode}$/XRT, and (i) is from $\textit{GOES}$/SXI.}
\label{fig:pre_eruptive}
\end{figure*}

\begin{figure*}[!ht]
\centering
\includegraphics[width=.9\textwidth,clip=]{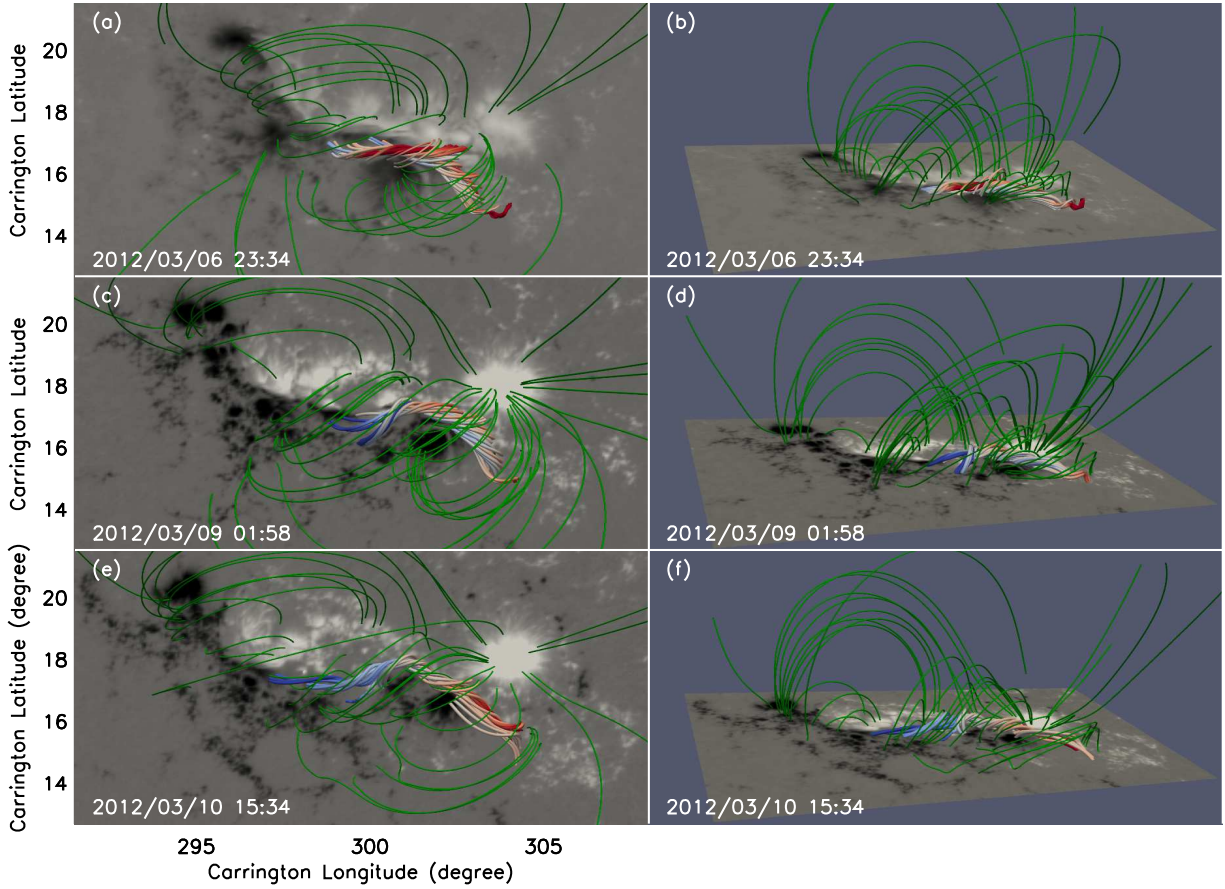}
\caption{Magnetic structure modeled by NLFFF extrapolation. Left (right) panels show the top (side) view of magnetic field lines rendered on the HMI LOS-magnetograms before each eruption. The sheared and twisted field lines, along the southwest polarity inversion line (SW-PIL), are shown in blue-red lines. The green lines show the overlying magnetic field in the corona.}
\label{fig:twisted}
\end{figure*}

On March 6 at 23:35 UT, $\sim$ one hour prior to the first solar eruption, we observed a coherent hot-channel structure (HCS) to lie along the SW-PIL (see Figure \ref{fig:pre_eruptive}(a)). Hot channels are so named as they are observed in hot passbands of the AIA i.e., 94 \AA~(6 MK) and 131 \AA~(10 MK) (e.g., \citealt{Zhang_etal_2012}). The temperature map, obtained through differential emission measure (DEM) method (e.g., \citealt{Cheng_etal_2012}), indicated that it was hotter than 7 MK. It was also observed as sigmoid in X-ray images (see Figure~\ref{fig:pre_eruptive}(c)). These are called sigmoid because of their forward-S or reverse-S shape (e.g., \citealt{Rust_and_Kumar_1996}). Both the HCS and X-ray sigmoids are considered as the observational proxy of MFR (e.g., \citealt{Green_etal_2007}; \citealt{Zhang_etal_2012}). NLFFF extrapolation results showed that long, sheared and twisted magnetic field lines were lying along the SW-PIL (see the top panels of Figure~\ref{fig:twisted}). Both observational and extrapolation results suggest the existence of an MFR in the SW sub-region of the AR. As discussed in section~\ref{eruption}, the HCS was observed to erupt on March 7 around 01:05 UT. A coronal structure was seen to reform by March 9 01:30 UT, in the SW sub-region over the period of two days. Its morphology in EUV and X-ray images was similar to the coronal structure observed on March 6 (see the middle panels of Figure~\ref{fig:pre_eruptive}). Ensuing the eruption on March 9 03:00 UT, another coronal structure of identical morphology (see the lower panels of Figure~\ref{fig:pre_eruptive}) was observed to reform in the period of one and half days. The extrapolation results showed that the coronal structure (on both days) was sheared and twisted like the observed structure on March 6 (see Figure~\ref{fig:twisted}). This suggests that MFR of identical shape was reformed before each homologous eruption. 

A persistent convergence and flux cancellation were observed in the SW sub-region. Also, there was a continuous shearing motion between the opposite magnetic fluxes. Shearing motion and flux cancellation are well known to generate long and twisted field lines along the PILs (e.g., \citealt{van_Ballegooijen_etal_1989}). This suggests that the shearing motion and flux cancellation were responsible for the recurrent formation of MFR along the SW-PIL.

\begin{figure*}[!ht]
\centering
\includegraphics[width=.9\textwidth,clip=]{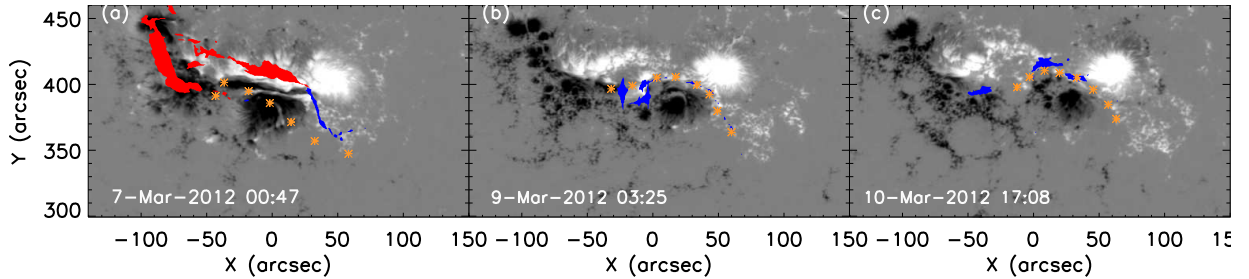}
\caption{Confined flare during the precursor phase of each eruption. The contours of flare ribbons during the precursor phase activity, as observed in AIA 1600 \AA, are overplotted on the HMI LOS-magnetogram in blue. The red contours in (a) show the flare ribbons from an earlier eruption. The orange asterisks indicate the outline of the hot-channel structure observed in AIA 131 \AA.}
\label{fig:precursor_ribbon}
\end{figure*}

\begin{figure*}[!ht]
\centering
\includegraphics[width=.7\textwidth,clip=]{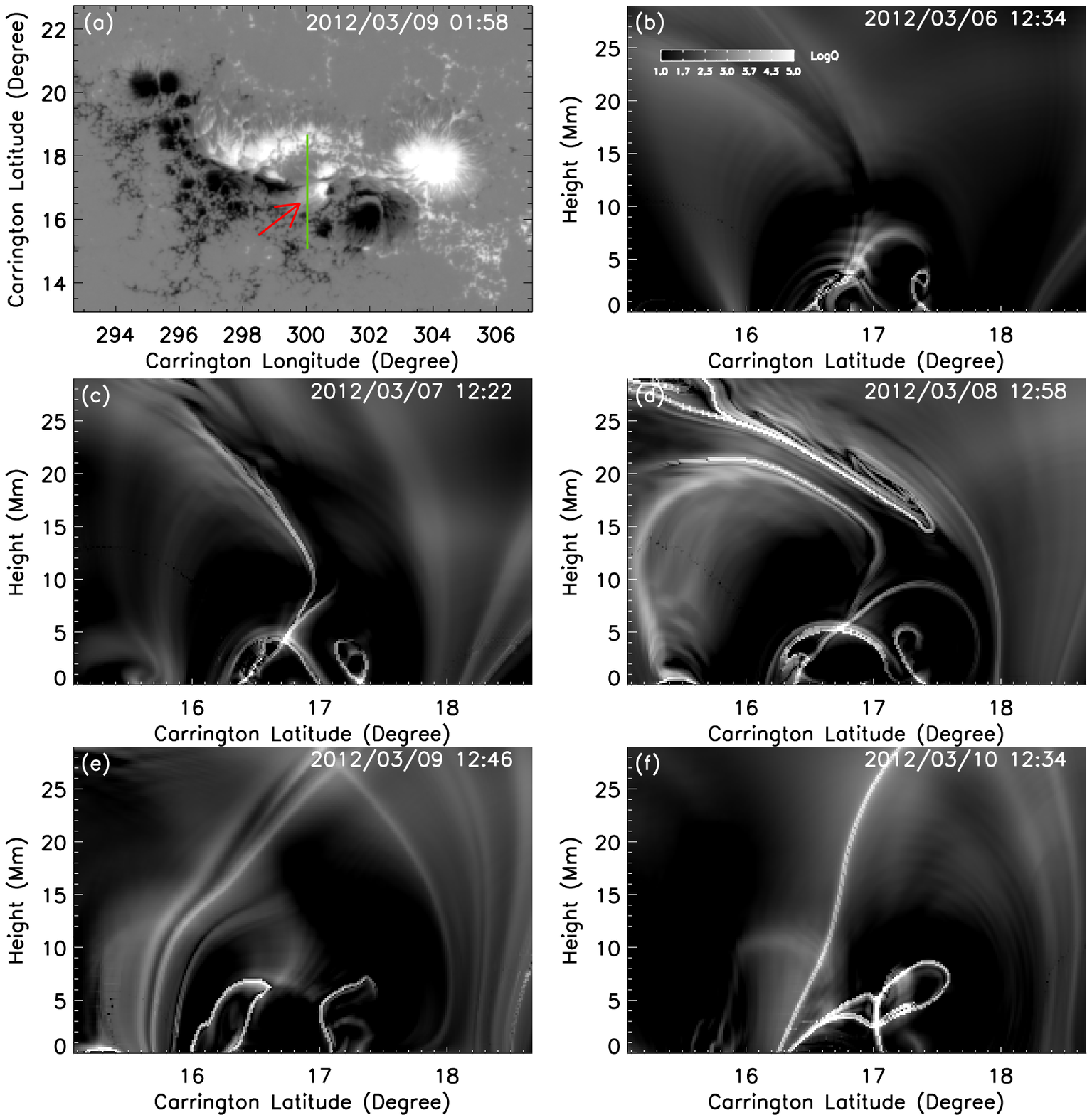}
\caption{The magnetic topology at different evolutionary epochs of the AR 11429. (a) HMI LOS-magnetogram on March 9, red arrow shows the area where positive magnetic flux is almost surrounded by negative flux. (b-f) show the vertical slice of logarithm squashing factor Q (Log(Q) $>$ 5) at the location of the green line in (a). Left panels show similar magnetic topology before each eruption.}
\label{fig:q_map}
\end{figure*}

\subsection{\textbf{Initiation of the eruption}}
A confined flare was observed on the left side of the SW sub-region during the PP of each eruption. It was observed as a small bump in the SXR intensity profile, a brightening of HCS, and the appearance of flare ribbons (see Figure \ref{fig:precursor_ribbon} for the flare ribbons). The SXR peak of the PP for the first eruption was not observed due to the contamination from another nearby event (as discussed in \ref{eruption}; see Figure~\ref{fig:dimming}), nevertheless other signatures were observed. Identical PP could suggest that the triggering mechanisms were the same for the three solar eruptions (e.g., \citealt{Vemareddy_2017}). The confined flare was observed near the small chunk of positive magnetic flux (positive-chunk), which was almost surrounded by the negative magnetic flux (shown by a red arrow in Figure~\ref{fig:q_map} (a)). Using NLFFF extrapolation results and codes developed by R. Liu and J. Chen (See \citealt{Liu_etal_2016}), we calculated the squashing factor Q in the AR 11429. The contours of high-Q are considered as the proxy of quasi-separatrix layers (QSLs; \citealt{Titov_etal_2002}). Flare ribbons appeared around the positive-chunk, therefore we analyzed the magnetic topology around it. The vertical-slice of Q-map in the middle of it revealed the presence of semicircular QSLs (see Figure \ref{fig:q_map}), extending from one end to the other end of it. The previous study by \cite{Polito_etal_2017} of the AR 11429, on March 9, had found 3D dome-shaped QSLs surrounding the positive-chunk. Here, the vertical slice of dome-shaped QSLs is seen as semicircular QSLs. The semicircular shape of QSLs was maintained before each eruption, though its size changed over the period. QSLs are the probable sites of magnetic reconnection (\citealt{Demoulin_etal_1996, Demoulin_etal_1997}). The whole positive-chunk was very dynamic throughout the evolution of the AR. Also, new flux emergence was observed there before March 9. Therefore, the shearing motion or flux emergence could have triggered the magnetic reconnection there. 

\begin{figure*}[!ht]
\centering
\includegraphics[width=.9\textwidth,clip=]{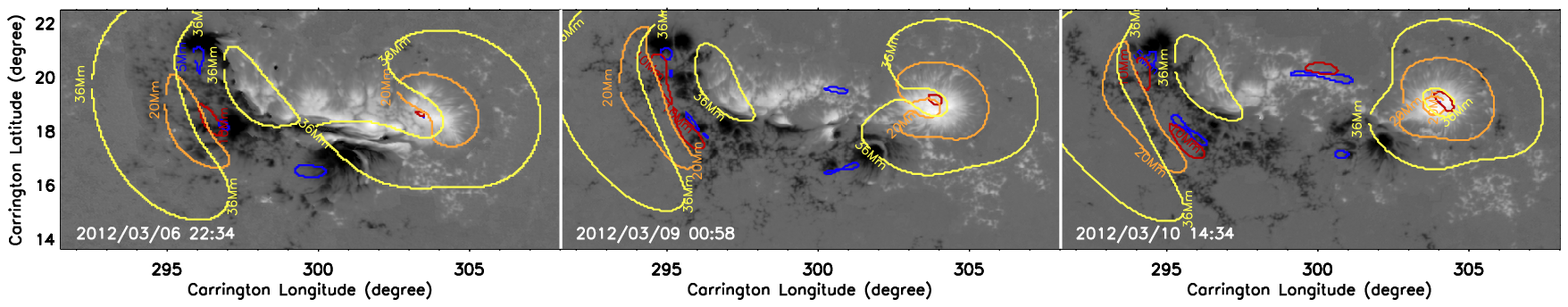}
\caption{Contours of critical decay index ($n \geq 1.5$) for the AR 11429. The contours are overplotted on HMI LOS-magnetogram and different colors represent the heights at which decay index is calculated.}
\label{fig:current_decay_map}
\end{figure*}

The MFR, along SW-PIL, was situated between two big magnetic poles and thus a strong overlying magnetic strapping force can be inferred above it. Strong overlying fields can inhibit successful eruption from ARs (e.g., \citealt{Sun_etal_2015}). The rate at which the strength of overlying magnetic fields decreases with height can be analyzed using the decay-index ($\textit{n}$). An MFR becomes torus-unstable if $\textit{n}$ is greater or equal to the critical value ($\textit{n}_{crit}$ = 1.5; \citealt{Bateman_1978}; \citealt{Kliem_and_Torok_2006}). Using NLFFF results we calculated the decay-index at different height as,
\begin{equation}
n = \frac{-d log {B}_{h}}{dlogz}
\end{equation}
Here, $B_{h}$ is the horizontal component of the magnetic field, and $\textit{z}$ is the radial distance from the solar surface. Figure~\ref{fig:current_decay_map} shows the distribution of super-critical decay-index ($\textit{n}$ $\ge$ 1.5) at different heights before each eruptions. The closed contours represent the area where horizontal magnetic fields are very weak. An erupting MFR would preferably pass through such contours (e.g., \citealt{Chintzoglou_etal_2015}). In the SW sub-region, the contours changed and moved more toward the positive polarity with evolution. This could be the reason for the observed change in the direction of successive CMEs, as discussed in section~\ref{eruption}. Also, it is clear from the decay-index maps that it would have been hard for low-lying MFR to become torus-unstable and erupt from the SW-PIL. Therefore, it is most probable that the PP flare facilitated the eruption of underlying MFR by weakening the constraint of overlying magnetic field.

\section{\textbf{Discussion and Conclusion}}
\label{DC}
Our primary goal of the present study is to understand the recurrent homologous solar eruptions from the AR 11429. There were three homologous eruptions from the southwest (SW) sub-region within 87 hours. The first eruption was associated with X-class flare and the other two eruptions were associated with M-class flares. Nevertheless, the shape of X-ray profiles of the three flares was very similar. The same photospheric location, similar evolution and location of flare ribbons, EUV dimmings, and identical coronographic morphology of the CME invoked them as homologous eruptions. 

Our main result of the study is that the continuous flux cancellation due to shearing motions and the convergence of opposite magnetic bipoles along the PILs was the primary factor for the homologous solar eruptions. Flux emergence was observed only in the initial evolutionary period of the AR, flux cancellation was observed throughout the evolution that span all three events, including prior to the first event. While flux emergence was important and relevant in AR 11429, our study suggests that long-term flux cancellation and shearing motions were the most probable mechanisms that produced homologous eruptions. The photospheric motions of magnetic flux injected magnetic helicity and magnetic energy in the corona gradually. Stored helicity and energy released intermittently during solar eruptions. We believe that due to continuous shearing and converging motions of the magnetic flux, there was a  cycle of storage and release of magnetic helicity and magnetic energy from the AR. In general, opposite magnetic poles of an emerging bipoles (conjugate pair) move away from each other, the separation distance depends on the flux content of the emerged bipoles (\citealt{Wang_and_Sheeley_1989}). In the present case, we did not observe the separation between opposite poles in SW sub-region, instead they were converging towards each other throughout the observational period. Therefore, we believe that these were non-conjugate pair. 
It was accompanied by the persistent shearing motion and magnetic flux cancellation.~\cite{Chintzoglou_etal_2019} suggests that the convergence of opposite magnetic poles of non-conjugated pair leads to the magnetic flux cancellation and shearing motions, which could produce solar eruptions. 

A filament was lying along the SW-PIL throughout the observational period and was observed soon after the eruptions. A filament survives and would be visible after a partial eruption. Partial eruption could happen either by the splitting of a single flux rope during the eruption(e.g., \citealt{Gibson_and_Fan_2006}; \citealt{Zhang_etal_2015}) or by the eruption of upper part of double-decker system (e.g., \citealt{Liu_etal_2012}; \citealt{Kliem_etal_2014}). Though a double-decker system was observed along the SW-PIL before the eruption on March 10 (see \citealt{Dhakal_etal_2018}), it is hard to say whether a single flux rope split or a double-decker system was present before the eruption on March 7 and March 9. Nevertheless, we believe that there was a partial eruption of the magnetic system from the SW-PIL. 

A magnetic structure was formed along the SW-PIL before each eruption. It was observed as a coherent hot-channel structure in AIA EUV and sigmoid in X-ray images. Reforming HCS/sigmoid was almost identical in appearance. HCS and sigmoid are considered as an observational proxy of magnetic flux rope (MFR). Long, sheared and twisted magnetic field lines along the SW-PIL were found in our extrapolation results. Therefore, we believe that an MFR was formed and existed before each eruption. \citealt{Chatterjee_and_Fan_2013} studied the emergence of twisted flux rope in magnetohydrodynamic simulation. They suggest that the partial eruption and reformation of MFR can produce homologous eruptions. In their simulation, reformed MFR was appeared as X-ray sigmoid. In the present case, flux emergence was going on before the first eruption. It is possible that the erupting MFR, before the first eruption, could have emerged from the sub-photosphere. However, since there were continuous shearing motion and flux cancellation in the SW sub-region, spanning the three eruptions. Also, after partial eruption an erupting magnetic structure of similar morphology was reformed. Therefore we believe that, most probably the shearing motion and flux cancellation were the primary mechanisms that formed the erupting MFR along the SW-PIL.

At last, we want to discuss about the identical precursor phase (PP) activity in the AR. It was identified as a confined flare surrounding an island of positive flux at the left side of the SW sub-region. A complex magnetic configuration, semicircular quasi-separatrix layers (QSLs), was inferred to exist in the region. The magnetic configuration was almost the same before each eruption, we believe this was due to the reformation of identical magnetic structures in the region. It is arguable that the identical PP was due to reconnection at identically shaped QSLs, either due to flux emergence or shifting of the island itself. As a strong overlying magnetic strapping force, due to overlying magnetic field, was inferred above the MFR lying along the SW-PIL. Our study suggests that the PP flare facilitated the eruption by removing the overlying magnetic strapping force above the MFR.

\acknowledgments
We thank the referee for the constructive comments and suggestions to improve the presentation of the results. HMI and AIA are instruments onboard \textit{SDO}, a mission for NASA's Living With a Star Program. S.D. is supported by the DKIST Ambassador program. Funding for the DKIST Ambassadors program provided by the National Solar Observatory, a facility of the National Science Foundation, operated under Cooperative Support Agreement number AST-1400405. J. Z. is supported by NASA grant NNH17ZDA001N-HSWO2R.  N. K. is supported by NASA grant NNX16AH87G S01.

\bibliography{homologous_eruption}
\bibliographystyle{aasjournal}

\end{document}